\documentclass{mn2e}
\usepackage{epsf}
\def\msun{{\rm M_{\odot}}}

\title[Accretion Rates and Beaming in ULXs]
{Accretion Rates and Beaming in Ultraluminous X--ray Sources}

\author[A. R. King] {A. R. King$^{1}$ \\
$^1$Theoretical Astrophysics Group, University of Leicester, Leicester
LE1 7RH}

\date{\today}

\volume{000}

\pagerange{\pageref{firstpage}--\pageref{lastpage}} \pubyear{2005}

\begin{document}

\label{firstpage}

\maketitle

\begin{abstract}
I show that extreme beaming factors $b$ are not needed to explain ULXs
as stellar--mass binaries. For neutron star accretors one typically
requires $b \sim 0.13$, and for black holes almost no beaming ($b \sim
0.8$). The main reason for the high apparent luminosity is the
logarithmic increase in the limiting luminosity for super--Eddington
accretion. The required accretion rates are explicable in terms of
thermal--timescale mass transfer from donor stars of mass $6 -
10\msun$, or possibly transient outbursts. Beaming factors $\la 0.1$
would be needed to explain luminosities significantly above
$10^{40}L_{40}$~erg\, s$^{-1}$, but these requirements are relaxed
somewhat if the accreting matter has low hydrogen content.

\end{abstract}


\section{Introduction}

The nature of the ultraluminous X--ray sources (ULXs) is still not
definitively settled. In one view they are intermediate--mass black
holes (IMBH) accreting at rates below their Eddington limits. In this
paper I shall adopt the contrary view (cf King et al., 2001) that they
represent a very bright and unusual phase of X--ray binary evolution,
in which the compact object is fed mass at a rate $\dot M$ well above
the usual Eddington rate $\dot M_E$. In this picture the large
apparent X--ray luminosity $L_X = 10^{40}L_{40}$~erg\, s$^{-1}$ is a
result of two effects (see below). First, the bolometric luminosity
genuinely exceeds the usual Eddington limit by a factor $\ln(\dot
M/\dot M_E)$, which can be significant. Second, the X--rays may be
collimated by a factor $b$ by scattering within an optically thick
biconical outflow. These conditions could in principle reflect a
genuine state of high mass transfer, or a transient outburst (King,
2002). Comparing this picture with observation is currently
complicated by the fact that there is freedom in choosing both the
super--Eddington factor $\dot M/\dot M_E$ and the beaming factor $b$.

In several ULXs it is possible to identify a soft X--ray component,
and indeed infer a lengthscale $R \sim 10^9$~cm associated with it. I
show here that in the context of the adopted model for ULXs, this
quantity and $L_X$ essentially fix both the Eddington and beaming
factors for a given mass $M_1$ of the accretor.

\section{Super--Eddington Accretion}

An accretor supplied with mass at a super--Eddington rate arranges to
expel matter from its accretion disc in such a way that it never
exceeds the local Eddington luminosity (Shakura \& Sunyaev, 1973; cf
Begelman, King \& Pringle, 2006; Poutanen et al., 2007). In this
picture there is a characteristic lengthscale $R_{\rm sph}$ where the
mass inflow first becomes locally Eddington (also called the
spherization or trapping radius) (cf Begelman et al., 2006, eq 14).
Following Shakura \& Sunyaev (1973) we write
\begin{equation}
R_{\rm sph} = {27\over 4}{\dot M \over \dot M_E}R_s,
\label{rsph}
\end{equation}
where $R_s = 2GM_1/c^2 = 3\times 10^5m_1$~cm is the Schwarzschild
radius of the accretor. 

The nature of the accretion flows outside and inside this radius
differ markedly. The region outside $R_{\rm sph}$ releases accretion
luminosity $\sim L_E$ in the usual way, where $L_E = 1.6\times
10^{38}m_1$ is the Eddington luminosity (for hydrogen--rich
material). But at and within $R_{\rm sph}$, radiation pressure drives
an outflow which keeps the local energy release very close to the
Eddington value and creates a biconical geometry, collimating the
outgoing radiation. This region releases a bolometric luminosity
\begin{equation}
L_{\rm acc} \simeq L_E\left[1 + \ln\left({\dot M\over\dot
M_E}\right)\right].
\label{lacc}
\end{equation}
Because of the geometric collimation by a factor $1/b \geq 1$, an
observer viewing such a disc in directions within one of the cones
sees an {\it apparent} bolometric luminosity
\begin{equation}
L \simeq {L_E\over b}\left[1 + \ln\left({\dot M\over\dot M_E}\right)\right]
\label{l}
\end{equation}
(cf Shakura \& Sunyaev, 1973; Begelman, King, \& Pringle, 2006). The
enhancement of $L$ over $L_E$ by beaming and the logarithmic factor
are the basis of the interpretation of ultraluminous X--ray sources
(ULXs) as hyperaccreting stellar--mass binaries (Begelman, King \&
Pringle, 2006; Poutanen et al., 2007).

Because the accretion flow within $R_{\rm sph}$ is optically thick, the
disc region there must produce a soft blackbody component (by
reprocessing of harder radiation, if for no other reason). This must
have characteristic lengthscale $R_{\rm sph}$ and luminosity $L_{bb}
\ga L_E$. This is only a part of the accretion luminosity
(\ref{lacc}), the harder power--law--like component characteristic of
ULXs resulting from some other process such as comptonization or
direct release of accretion energy. The observability of the soft
blackbody component depends on its temperature $T = (L_{bb}/4\pi\sigma
R_{\rm sph}^2)^{1/4}$, which for typical ULX parameters $L_{bb} \sim
10^{39} - 10^{40}$~erg\, s$^{-1}$, $R_{\rm sph} \sim 10^8 - 10^9$~cm
is always in the range $0.1 - 1$~keV. Several ULXs are observed to
show such components, and it is reasonable to infer that more would do
so if their temperatures were slightly higher or the photoelectric
absorption slightly lower.

This interpretation of the blackbody component implies that it should
be collimated by a similar beaming factor $b$ to the harder radiation,
and accordingly is likely to dominate the similar but unbeamed
component produced outside $R_{\rm sph}$. Accordingly we have to
consider the effect of beaming on this component.
  
Since the inferred blackbody temperature $T \sim 0.1$~keV
is determined by the spectral shape, and thus unaffected by
collimation, the observer infers a total blackbody luminosity
\begin{equation}
4\pi R^2\sigma T^4 = {4\pi\over b}R_{\rm sph}^2\sigma T^4
\end{equation}
from this component. Thus we have $R_{\rm sph} = b^{1/2}R$. From eqn
(\ref{rsph}) we find the Eddington ratio
\begin{equation}
{\dot M \over \dot M_E} = {4R_{\rm sph}\over 27R_s} = {490R_9
b^{1/2}\over m_1}.
\label{mdot}
\end{equation}
From eqn (\ref{l}) we deduce finally
\begin{equation}
b = {0.016m_1\over L_{40}}\left[1 + \ln\left({490R_9\over
m_1}b^{1/2}\right)\right]
\label{b}
\end{equation}

We note that this equation uses the assumption that the characteristic
lengthscale of the blackbody component is $R_{\rm sph}$, but does not
assume that this component necessarily dominates the bolometric
emission there.

Given a value of $m_1$, eqn (\ref{b}) is a transcendental equation for
$b$. One can show by standard methods that it always has two roots for
$b$ unless $m_1 $ is very large ($\ga 3300R_9^2/L_{40}$). For typical
ULX parameters one root is small, of order $10^{-4}$, while the other
is in the range $b \sim 0.1 - 1$. For the small root the logarithm on
the rhs is small, implying $\dot M/\dot M_E \sim 1$, while for the
root close to unity the logarithm is $O(1)$, so that $\dot M/\dot M_E
>> 1$.

Physically these roots correspond respectively to two different
possible cases. For the small root case, collimation dominates the
logarithmic increase in the total accretion luminosity above the
standard Eddington value, and is thus the main reason for the high
apparent luminosity of the ULX. Hence this case corresponds to an only
mildly super--Eddington flow nevertheless producing a highly
anisotropic radiation pattern. In the case where $b$ is close to
unity, collimation is weak, and the system appears bright largely
because of the logarithmic increase in the true accretion
luminosity. This latter case appears physically more plausible, and I
shall adopt the larger root for $b$ in what follows. I consider masses
$m_1 = 1.4, 10$ corresponding to a neutron--star and black hole
accretor respectively, and ask what values of $b$ and $\dot M$ these
require for two typical observed cases. (Note that all the formulae
above hold for accretion on to a neutron star, or indeed any other
star, provided only that it is sufficiently super--Eddington that its
physical radius is smaller than the spherization radius $R_{\rm sph}$
defined in (\ref{rsph}). Of course if some of the accretion luminosity
is emitted from the surface of the star, this could modify the
relation between accretion rate and luminosity and slightly change the
factor 27/4 in equation (\ref{rsph}).)

\subsection{Typical ULX, $L_{40} = R_9 = 1$} 

Taking $\dot M_E = 2.5\times 10^{-8}m_1~\msun\, {\rm yr}^{-1}$,
corresponding to hydrogen--rich accretion with a radiative
efficiency of $0.1c^2$, we find consistent solutions with
\begin{equation}
m_1 = 1.4,\ \  b = 0.13, \ \ \dot M/\dot M_E = 125, \ \ \dot M =
3.1\times 10^{-6}\msun\, {\rm yr}^{-1}
\end{equation}

and

\begin{equation}
m_1 = 10,\ \  b = 0.76, \ \ \dot M/\dot M_E = 43, \ \ \dot M =
1.0\times 10^{-5}\msun\, {\rm yr}^{-1}.
\end{equation}

\subsection{Hydrogen--poor accretion}

If the accreting matter has little hydrogen, the change in mean mass
per electron alters (\ref{b}) to
\begin{equation}
b = {0.028m_1\over L_{40}}\left[1 + \ln\left({490R_9\over
m_1}b^{1/2}\right)\right]
\label{he}
\end{equation}
and the Eddington accretion rate for efficiency $0.1c^2$ becomes
$\dot M_E = 4.2\times 10^{-8}m_1~\msun\, {\rm
  yr}^{-1}$.

Then if a neutron--star ULX accretes such matter we find
\begin{equation}
m_1 = 1.4,\ \  b = 0.24, \ \ \dot M/\dot M_E = 171, \ \ \dot M =
1.0\times 10^{-5}\msun\, {\rm yr}^{-1}.
\end{equation}

A black hole ULX ($m_1 = 10$) with $R_9 = 1$ accreting this kind of matter 
actually produces a luminosity $L = 1.36\times 10^{40}$~erg\, s$^{-1}$
without any beaming at all (i.e. $b=1$) if it accretes with 
\begin{equation}
m_1 = 10,\ \  b = 1, \ \ \dot M/\dot M_E = 49, \ \ \dot M =
2.0\times 10^{-5}\msun\, {\rm yr}^{-1}.
\end{equation}

For a very bright ULX with $L_{40} = 10, R_9 = 1$ accreting
hydrogen--poor matter we find
\begin{equation}
m_1 = 10,\ \  b = 0.10, \ \ \dot M/\dot M_E = 16, \ \ \dot M =
6.7\times 10^{-6}\msun\, {\rm yr}^{-1}.
\end{equation}

\section{Discussion}

The simple calculations of subsections 2.1, 2.2 above illustrate the
following general points.

(a) extreme beaming factors are not needed to explain ULXs as
stellar--mass binaries. With neutron stars one typically requires $b
\sim 0.13$, and with black holes almost no beaming ($b \sim 0.8$)
except for the very brightest ULXs ($L_X = 10^{41}$~erg\,
s$^{-1}$). Here beaming approaches 10\%, since $\ln(\dot M/\dot M_E)$
cannot realistically exceed 10, and $L_E$ is similarly limited to a
few times $10^{39}$~erg\,s$^{-1}$.

(b) typical ULXs with $L_X = 10^{40}L_{40}$~erg\, s$^{-1}$ and
$R\simeq 10^9$~cm can be explained in either of two ways: 

\noindent
neutron star binaries accreting at $\sim 100$ times their Eddington
rates, i.e at $\dot M \sim 3\times 10^{-6}\msun\, {\rm yr}^{-1}$,

\noindent
or black hole binaries accreting at $\sim 40$ times their Eddington
rates, i.e. at $\dot M = 1.0\times 10^{-5}\msun\, {\rm yr}^{-1}$.

I note that both required accretion rates are explicable in terms of
thermal--timescale mass transfer. This occurs when a high-- or
intermediate--mass radiative star transfers mass to a less massive
companion, and gives rates $\dot M \sim 3\times
10^{-8}m_2^{2.6}\msun\, {\rm yr}^{-1}$, where $m_2$ is the companion
mass in $\msun$ (King \& Begelman, 1999). We thus require companion
masses $M_2 \ga 6\msun, 10\msun$ in the neutron--star and black--hole
cases respectively, with black--hole masses slightly greater than
$10\msun$. As emphasized by King et al. (2001), this type of binary
evolution predicts source lifetimes and numbers in good agreement with
observation. Some very bright transients may be able to achieve these
accretion rates during outbursts (cf King, 2002). This is likely to be
the only way of making ULXs in old stellar populations (King et al.,
1997).

(c) accretion of hydrogen--poor matter generally reduces the
requirements for beaming still further, and is the most likely
explanation for the very brightest ULXs ($L_X = 10^{41}L_{41}$~erg\,
s$^{-1}$). Accretion from massive Wolf--Rayet type companions is the
most likely origin of these very high luminosities, although other
explanations are possible in rare cases (cf King \& Dehnen, 2005).

\end{document}